# Exchange bias in van der Waals MnBi$_2$Te$_4$/Cr$_2$Ge$_2$Te$_6$ heterostructure

Jing-Zhi Fang,[1,2,5,*] Hao-Nan Cui,[2,4,*] Shuo Wang,[2,3,*] Jing-Di Lu, [6,*] Xin-Jie Liu[2,7], Guang-Yu Zhu[2], Mao-Sen Qin[2], Jian-Kun Wang[2], Ze-Nan Wu[2], Yan-Fei Wu[7], Shou-Guo Wang[7], Zhongming Wei,[1,5,†] Jinxing Zhang, [6,†] Ben-Chuan Lin,[2,3,†] Zhi-Min Liao[2,4,†] and Dapeng Yu[2,3]

[1]*State Key Laboratory of Superlattices and Microstructures, Institute of Semiconductors, Chinese Academy of Sciences, Beijing 100083, China*

[2]*Shenzhen Institute for Quantum Science and Engineering, Southern University of Science and Technology, Shenzhen, 518055, China*

[3]*Guangdong Provincial Key Laboratory of Quantum Science and Engineering, Southern University of Science and Technology, Shenzhen, 518055, China*

[4]*State Key Laboratory for Mesoscopic Physics and Frontiers Science Center for Nano-optoelectronics, School of Physics, Peking University, Beijing 100871, China*

[5]*Center of Materials Science and Optoelectronics Engineering, University of Chinese Academy of Sciences, Beijing 100049, China*

[6]*Department of Physics, Beijing Normal University, Beijing 100875, China*

[7]*Beijing Advanced Innovation Center for Materials Genome Engineering, School of Materials Science and Engineering, University of Science and Technology Beijing, Beijing 100083, China*

[*] *These authors contributed equally to this work.*

[†]Corresponding author. Email: zmwei@semi.ac.cn; jxzhang@bnu.edu.cn; linbc@sustech.edu.cn; liaozm@pku.edu.cn

**Abstract:**

The layered van der Waals (vdW) material MnBi$_2$Te$_4$ is an intrinsic magnetic topological insulator with various topological phases such as quantum anomalous Hall effect (QAHE) and axion states. However, both the zero-field and high-temperature QAHE are not easy to realize. It is theoretically proposed that the exchange bias can be introduced in the MnBi$_2$Te$_4$/ferromagnetic (FM) insulator heterostructures and thus opens the surface states gap, making it easier to realize the zero-field or high-temperature QAHE. Here we report the electrically tunable exchange bias in the van der Waals MnBi$_2$Te$_4$/Cr$_2$Ge$_2$Te$_6$ heterostructure. The exchange bias emerges over a critical magnetic field and reaches the maximum value near the magnetic band gap. Moreover, the exchange bias was experienced by the antiferromagnetic (AFM) MnBi$_2$Te$_4$ layer rather than the FM layer. Such van der Waals heterostructure provides a promising platform to study the novel exchange bias effect and explore the possible high-temperature QAHE.

## Introduction

Magnetic loop hysteresis with single symmetric coercive field and saturation magnetization is a basic property in typical magnetic materials. In some cases, this intrinsic symmetry can be broken by introducing different phase systems to generate exchange coupling force, namely exchange bias(*1*). In general, an asymmetric loop shift along the external-field axis is observed in the FM/AFM heterostructure because of magnetic moment's pinning effect in two phases' boundary. Such loop offset, namely exchange bias, can be programmed and be used to commercial magnetic recording and related spintronic devices(*2*). In some magnetic materials, the exchange bias can also act as an indication of novel phases such as spin glass(*3, 4*) or other multi-magnetic phases(*5, 6*). Though the exchange-bias effect has been widely observed in metal-based magnetic heterostructures, it is still challenging to be realized in magnetic vdW heterostructures. Because the electron exchange is generally not expected in magnetic vdW heterostructures(*6*), and thus the absence of exchange bias is reasonable.

The magnetic topological insulator $MnBi_2Te_4$ (MBT) consisting of Te-Bi-Te-Mn-Te-Bi-Te septuple layers (SLs) by van der Waals interaction was predicted as an ideal intrinsic topological non-trivial material with a Neel temperature of 24 K(*7–9*). Bi-Te related topological properties and Mn-3d magnetism arranged by FM-intralayer/AFM-interlayer in c-axis provide abundant topological phases to explore, such as axion insulators and Weyl semimetals(*10–12*). Though the zero field QAHE has been reported (*9*), most groups only observed the quantized plateau at a much higher magnetic field, i.e. 6 T (*11, 13–15*), which restricts its promising application in future low-consumption electronics or topological quantum computation. The reason lies in the fact that the surface states of MBT is nearly gapless as revealed by the ARPES(*16–18*), indicating the absence of out-of-plane surface magnetization. In light of this situation, it is theoretically proposed that the surface magnetism can be induced by a ferromagnetic insulator and a relatively large exchange bias was predicted due to the ferromagnetic coupling between the AFM MBT and FM insulator layer(*13*). The strong exchange coupling stems from the strongly overlapped orbitals at the interface. In the meantime, the ferromagnetic insulator has little influence on the band structure of the MBT, which makes it possible to retain the topological properties of MBT and realize the zero-field or even high-temperature QAHE more easily.

In this work, we observed the electrically tunable exchange bias in van der WaalsMnBi$_2$Te$_4$/Cr$_2$Ge$_2$Te$_6$ heterostructure. The exchange bias was experienced by the antiferromagnetic (AFM) MnBi$_2$Te$_4$ layer and observed over a critical magnetic field, revealing the relation with the magnetic history. The shift of the hysteresis loop demonstrates the ferromagnetic coupling at the interface. What's more, it reaches a maximum value near the magnetic band gap, demonstrating an electrically tunable exchange bias effect. The exchange bias remains stable when changing different Hall measurement modes, ruling out possible local trivial disorder related effect. The nonlocal measurements, the magnetic force microscopy (MFM) experiments, and the control experiments were further conducted to help confirm the role of the magnetic proximity effect of the ferromagnetic insulator Cr$_2$Ge$_2$Te$_6$.

## Results

**Device configuration and the exchange bias**

The MBT flake was exfoliated by scotch tape and topped with Cr/Au electrodes by electron beam lithography (EBL) into standard Hall-bar structure. With dry transfer technique, we made Cr$_2$Ge$_2$Te$_6$ (CGT) onto the surface of MBT to form magnetic heterostructure. The sample thickness is $d$ = 9.8 nm, corresponding to 7-layer MBT. Fig. 1(A) marks six terminals as number ① to ⑥. The Hall resistance is $R_{41,53} = V_{53}/I_{41}$, where a current $I_{41}$ is applied between terminal ④ and ①, then the Hall voltage $V_{53}$ is measured. As shown in Fig. S1(A)-(B), the sample is located in the hole-carriers region with $T_N$ = 20 K and the measurement data unfolds higher quality loop in $V_g$ = 60 V. Firstly, we magnetized sample with a relatively high magnetic field to effectively align the moments and then repeatedly sweeped small-field-hysteresis curve, named PMP (positive magnetism polarized) and NMP (negative magnetism polarized), respectively.

For example, Fig. 1(B) exhibits negative exchange bias after magnetizing sample with a high magnetic field in 5 K. Specifically speaking, the sample was polarized in 2 T at first and then measured with ±1 T in PMP process. NMP process is the same procedure except the polarization in -2 T. The complete process is shown in Fig. S2. Because of strong coupling between longitudinal and transverse direction, the measurement results all have little vertical shift. But electrode asymmetry has nothing to do with moment-flip-related coercive field and the exchange bias. The positive and the negative sweep protocol both show that the exchange-bias field $B_{\text{EB}}$ (defined as $|\frac{Bc1(+/-)+Bc2(-/+)}{2}|$) is about 0.1 T. $B_{c1}$ ($B_{c2}$) is the smaller (bigger) coercive field showing specific transformation in Fig. 1(C,D). By changing negative (positive) polarized fields from -1 T (1 T) to -2 T (2 T) in 5 K, NMP and PMP process both clearly unveils two types of coercive fields when increasing external polarization fields. When polarized field is smaller than ±1.8 T, only $B_{c1}$ exists and the magnetic loops are symmetric as normal. The extraordinary loop shift emerges with the coexistence of $B_{c1}$ and $B_{c2}$ in larger NMP/PMP process, which is similar to the exchange bias in FM/AFM hetero-structure within topological insulator(*19, 20*). Besides, two Hall channels were measured simultaneously in order to exclude the influence of local effect in Fig. S3, in which case small multi-domain structure and trapped carriers can cause asymmetric behavior such as irreversible domain moving by Barkhausen effect(*21, 22*). Meanwhile, 4 T and 6 T loops' offset amplitude is the same and the training effect was not observed when sweeping loops repeatedly in 40 mK, as shown in Fig. S4 and S5. Therefore, the exchange-bias field $B_{\text{EB}}$ is not affected by changing maximum magnetic states and stable enough. The MFM observations help to further confirm the magnetization change in this process (Fig. S6).

**Electrically tunable exchange bias**

Furthermore, the exchange bias can be tuned by the back gate voltage $V_g$, namely the Fermi level. Fig. 2(A) shows the gate-voltage dependence of the exchange bias. The slope is negative because the Hall signal was measured as $R_{41,62}$ in the hole region as to

Fig. S1(A). When increasing $V_g$ to shift Fermi level close to the MBT's magnetic band gap, the contribution of ordinary Hall declines but the exchange-bias feature is almost fixed, further excluding its trivial origins. Moreover, the exchange bias shows an evident electrically tunable behavior with the maximum near the magnetic band gap (Fig.2(B)), which is a clear sign of the tunable exchange coupling strength between the ferromagnetic CGT layer and the antiferromagnetic MBT vdW layer. To clearly reveal the variation of exchange-bias process, the two vital parameters $B_{EB}$ and $B_c$ (defined as $|\frac{Bc1(+/-)-Bc2(-/+)}{2}|$) are extracted after subtracting linear background in Fig. 2(B). The complete double coercive states dependence of voltage within ±1 T and ±3 T is displayed in Fig. S7. We also find the slope of linear background is not monotonic when tuning Fermi level closer to magnetic gap, as showed in Fig. 2(C). The Hall coefficient is always positive because of hole-properties and the peak represents beginning of hole-to-electron transition, which indicates the Fermi level is quite near to the magnetic band gap.

The current/voltage ports were exchanged and three channels of Hall signals were measured simultaneously in the Fig. S8, verifying the Onsager relation. To clearly show $B_{c1}$ and $B_{c2}$ independently, we choose sweeping loop ranges ±1 T and ±3 T, which demonstrates this novel moment-flip relation spreads all over the sample and rule out the possible effects of trivial surface magnetic disorder in local areas.

Fig. 3(A) shows the temperature dependence of the exchange bias from 5 K to 20 K, which is close to MBT's intrinsic transition point. The anomalous Hall signal declines with rising temperature, which shows intrinsic moment evolution in Fig. 3(B), where $R_{sat}=|\frac{R_{sweep\ up(B=0)}-R_{sweep\ down(B=0)}}{2}|$. By calculating $B_{EB}$ and $B_c$ in Fig. 3(C), the exchange bias diminishes with increasing temperature.

**Nonlocal transport**

Nonlocal transport is always used to detect non-dissipative chiral edge states or other peculiar conducting channels in magnetic topological insulators. The contribution of asymmetric transport in non-chiral dissipative edge channels and in-plane magnetism in side surfaces might induce different coercive fields during the moment switching process(*23*). The quasihelical edge states and finite-size effect is found in other magnetic topological insulators by the specific kinks in $R_{xx}$ and $R_{xy}$ (*24*). To eliminate the influence of edge and side surface, we grounded three ports (terminals ②④⑥) in order to only allow bulk and top/bottom surface channels pass the sample, as shown in Fig. S9. $B_{c1}$ and $B_{c2}$ still exist with the same sweeping ranges ±1 T and ±3 T, which means its origin is confined to abnormal surface or bulk magnetic transport. This behavior can be further demonstrated in the nonlocal measurement shown in Fig. 4. When exchange $B_{c1}$ state and $B_{c2}$ state by adjusting different sweeping ranges, the channels of nonlocal signal have two totally different moment-flip-related behaviors, the hump for $B_{c1}$ and another dip for $B_{c2}$. All the signals are measured concurrently to ensure the relevance between nonlocal and Hall channels. Two types of nonlocal channels were further conducted, displaying consistent asymmetric hump characteristic in Fig. S10.

## Discussion

The surface magnetism of MBT can be magnetically proximitized by the FM insulator, without destroying the electronic band structure of the MBT film(*13*). Here the CGT stably pins the FM order of the proximitized layer of the MBT film, acting as an effective exchange bias. At the interface, the FM coupling would dominate in the MBT/FM insulator heterostructure(*13*), revealed by the observed negative exchange bias.The exchange bias effect in FM/AFM interface occurs when $K_{AFM} \cdot t_{AFM} > J_{ex}$, where $K_{AFM}$ and $t_{AFM}$ are anisotropy constant and thickness of AFM respectively, and $J_{ex}$ is the exchange coupling constant between FM/AFM(*25*). As the Fermi level is shifted to the magnetic band gap, the anisotropy energy of MBT would dominate over the interfacial exchange coupling and thus induce more evident exchange bias.

It is also noteworthy to point out that the exchange bias occurs only when the magnetic field is first polarized above a critical value. The pinning process observed by the magnetic force microscope (MFM) in Fig. S6 helps to clarify this process. The scanning area is marked by black frame in Fig. 1(A). After the NMP of -1 T, the part of CGT is reversed firstly and then the other part of MBT is changed. On the contrary, the NMP of -3 T provides a totally opposite result, then it can be deduced that the CGT and the covered MBT parts are pinned altogether. MFM picture helps divide two different parts of MBT, uncovered/covered topped CGT. Corresponding to transport procedure, in small sweeping ranges (±1 T), these beneath CGT's FM/AFM domain-related moments were not pinned together and only intrinsic magnetization reversal plays a role in $B_{c1}$ state. This symmetry overturn process can be broken by magnetizing with higher unilateral fields and the pinning effect at the FM/AFM interface emerges, which causes coexistence of $B_{c1}$ and $B_{c2}$ with exchange bias. Considering MBT's AFM coupling feature among interlayers, this negative loop shift does not originate from aligned intrinsic or extrinsic moments in the same direction between adjacent MBT layers, which will show positive loop moving when constantly rising external fields in field-cooling (FC) process because of the dominant AFM coupling dominant(*26, 27*), inconsistent with our observations in Fig. S4 and S5.

The nonlocal transport can also be explained using the proximity of the CGT layer. Multi-domain structure induced by the CGT's proximity can provide extra channels such as quantum percolation within chiral mode network(*28*) for nonlocal signals, which might also cause loop-shape larger nonlocal transport behavior. Besides, the hump-like peak corresponding to $B_{c1}$ state in small sweeping ranges can be connected with asymmetric magnetoresistance by considering carrier gyratory motion in the boundary of domain wall(*29–31*). Fig. S10 shows two types of nonlocal channel displaying asymmetric hump characteristic, confirming this extra field circulation.

The exchange bias observed in the heterostructure is experienced by the AFM MBT layer rather than traditional FM layer. Such novel effect comes from the special proximity effect in the vdW heterostructure as predicted by the theory(*13*), in which the FM insulator helps to pin the surface magnetic order while weakly disturb the electronic states of the MBT system. The control experiment of pure MBT (Fig. S11) shows no exchange bias in ±1 T or ±9 T sweeping ranges, indicating that CGT provides irreplaceable proximity effect and induce the exchange bias in the heterostructure.

The observation of the exchange bias in the MnBi$_2$Te$_4$/Cr$_2$Ge$_2$Te$_6$ helps corroborate that the surface magnetism can be introduced into the intrinsic magnetic topological insulator, paving the way to stable zero-field or even high-temperature QAHE. In light of the ongoing difficulties of realizing stable zero-field QAHE, future sandwich-structure experiments would be further conducted to introduce more exchange coupling realize stable QAHE.

In summary, we have reported the tunable exchange bias in the vdW heterostructure MnBi$_2$Te$_4$/Cr$_2$Ge$_2$Te$_6$. The stable exchange bias can be easily tuned by the gate voltage and magnetic field. Our observations corroborate the theoretical prediction of the exchange bias in the magnetic topological insulator/ferromagnetic insulator heterostructure. This work provides a new framework on exploring both novel exchange bias experienced by the antiferromagnetic magnetic topological insulators and high-temperature quantum anomalous Hall effect.

## Materials and Methods

### Devices fabrication

MBT nanosheets were prepared on $SiO_2$/Si substrates by mechanical exfoliation method in a glove box with $O_2$ and $H_2O$ contents less than 0.01 ppm. The Si substrate was covered with 285 nm $SiO_2$, so the substrate can also be used as the back gate. Before heterostructure establishment, a sharp needle was used to remove the thick sheet around the thin MBT. Then, the MBT is covered with a layer of CGT by dry transfer method. After that, the hall-bar pattern was prepared by standard electron beam lithography, and the Cr/Au electrodes was evaporated by electron beam evaporation. Almost all the processes are done in the high-vacuum circumstance. Finally, a layer of h-BN is transferred to cover the device as protection to reduce air pollution.

### Transport measurement

The electrical transport measurement was conducted in an Oxford cryogenic refrigerator. The lock-in amplifier was used to apply a constant AC current of 100 nA and measure the voltage simultaneously. The magnetic field was applied along the out-of-plane direction of the MBT films.

## Acknowledgments:

**Funding**

This work was supported by National Natural Science Foundation of China (Grants No. 12004158, No. 12074162, No. 91964201, No. 61825401, No. 62125404 and No. 11774004), National Key Research and Development Program of China (Grants No. 2020YFA0309300, No. 2018YFA0703703 and No. 2020YFB1506400), the Key-Area Research and Development Program of Guangdong Province (No. 2018B030327001) , Guangdong Provincial Key Laboratory (No. 2019B121203002), Hefei National Laboratory (2021ZD0303001) and the Strategic Priority Research Program of Chinese Academy of Sciences (Grant No. XDB43000000).

**Contributions:** Z.-M.L., B.-C.L., Z.-M.W. and D.Y. supervised the project. J.-Z.F., S.W. fabricated the sample and did the sample characterization with the necessary help from X.-J.L., G.-Y.Z., M.-S.Q., J.-K.W., Z.-N.W., Y.-F.W., S-G.W. and other authors. J.-Z.F., B.-C.L., S.W. and H.-N.C. did the quantum transport measurements, J.-D.L. and J.-X.Z. did the MFM experiments. H.-N.C., B.-C.L., J.-Z.F. analyzed the data and wrote the manuscript with the inputs from all authors.

**Competing interests:** The authors declare that they have no competing interests.

**Data and materials availability:** All data needed to evaluate the conclusions in the paper are present in the paper and/or the Supplementary Materials. Additional data related to this paper may be requested from the authors.

# Figures:

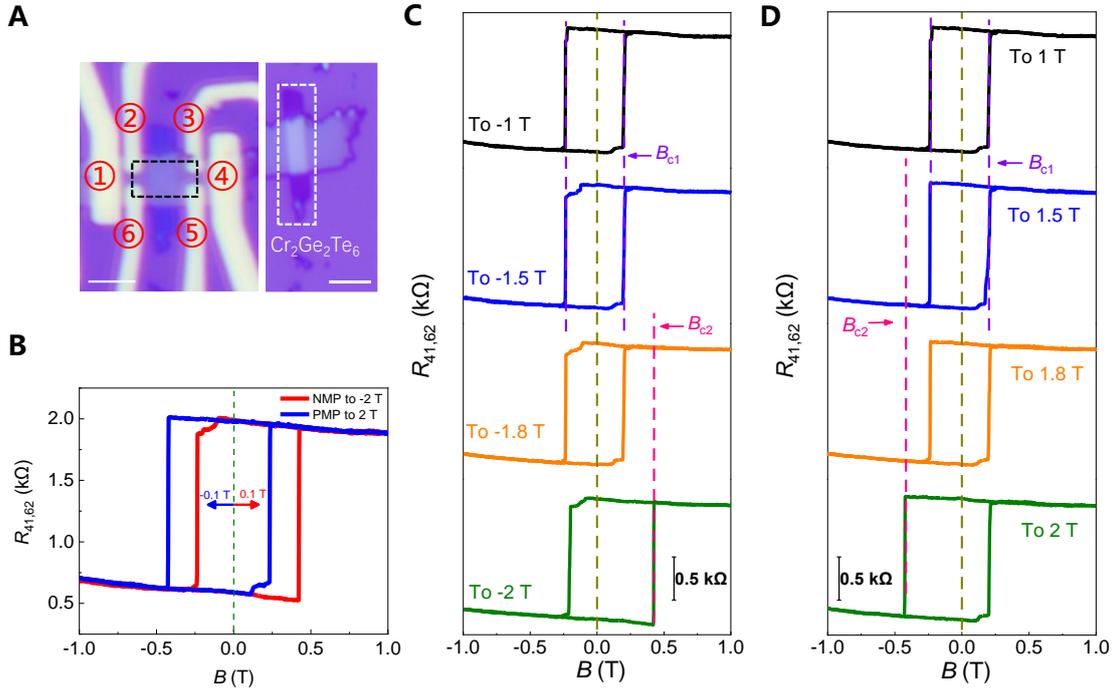

**Fig. 1. The optical image of the device and the observed exchange bias.** (**A**) The optical image of MnBi$_2$Te$_4$/Cr$_2$Ge$_2$Te$_6$ Hall-bar device. Six terminals are marked respectively from ① to ⑥. The orange dashed frame shows the topped Cr$_2$Ge$_2$Te$_6$. The scanning area of magnetic force microscope is marked by black lines. (**B**) Typical negative exchange bias after positive magnetism polarized (PMP) and negative magnetism polarized (NMP) process (2 T) in 5 K. Both polarized directions show symmetric $B_{EB}$ = 0.1 T. $B$ = 0 T is marked by green dashed line. (**C**) Increasing maximum negative polarized fields from -1 T to -2 T in 5 K. Minor (major) coercive field $B_{c1}$ ($B_{c2}$) is marked with purple (pink) lines. (**D**) is the same as (**C**) except for positive polarized fields from 1 T to 2 T. All the above measurement is in $V_g$ = 60 V.

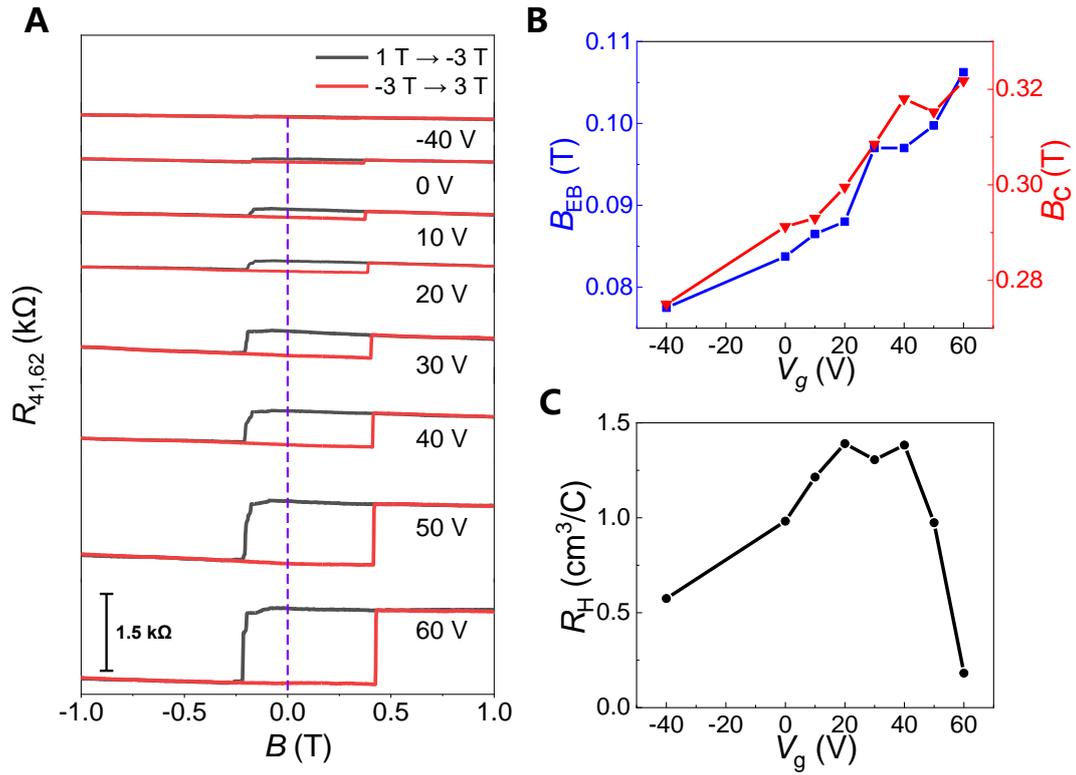

**Fig. 2. The gate-voltage dependence of the exchange bias.** (**A**) The gate-voltage dependence of exchange-bias with NMP process to -3 T from -40 V to 60 V in 5 K. The purple vertical line marks the position of zero point. (**B**) The back-gate dependence of exchange-bias field $B_{EB}$ and loop coercive field $B_c$ extracted from (**A**) after subtracting linear background. (**C**) Hall coefficient is extracted by ordinary Hall component. $R_H > 0$ indicates its hole-transport feature.

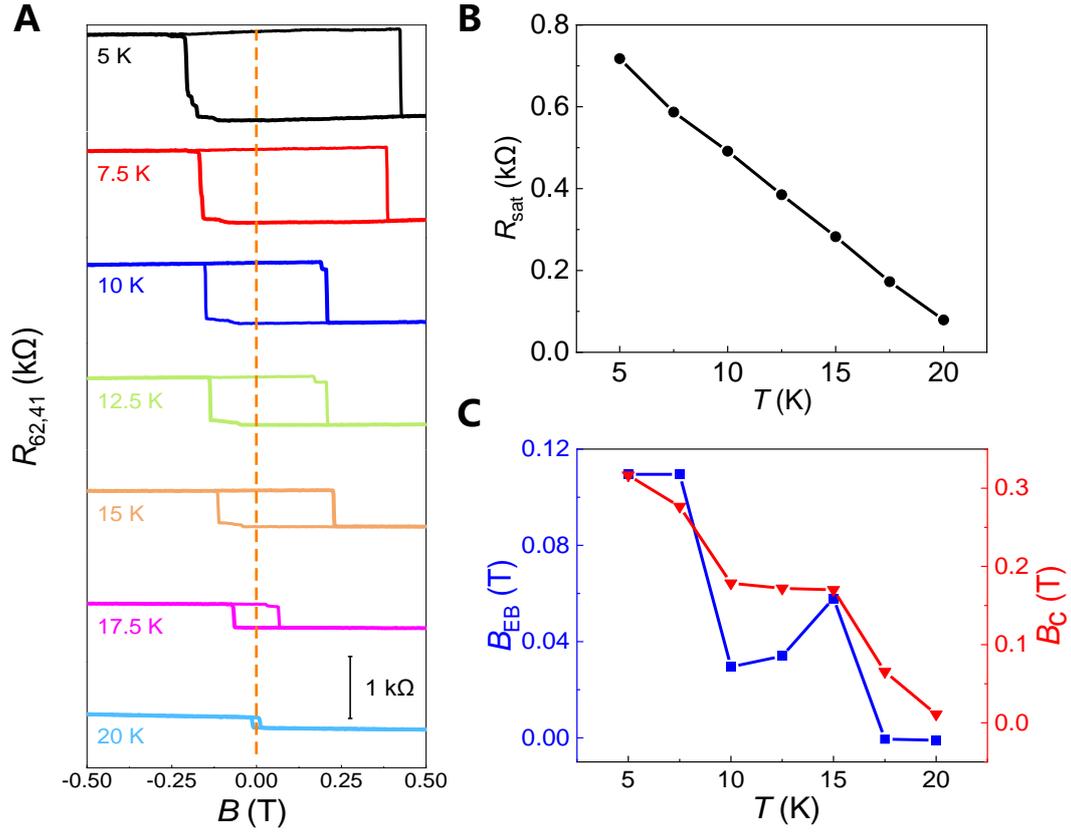

**Fig. 3. The temperature dependence of exchange bias.** (**A**) The temperature evolution of exchange bias from 5 K to 20 K at $V_g$ = 60 V. The orange vertical line marks the zero field. (**B**) The saturation resistivity of anomalous Hall signals at $B$=0 T with $R_{sat}=|\frac{Rsweep\ up(B=0)-Rsweep\ down(B=0)}{2}|$. (**C**) The temperature dependence of extracted exchange-bias field $B_{EB}$ and loop coercive field $B_c$.

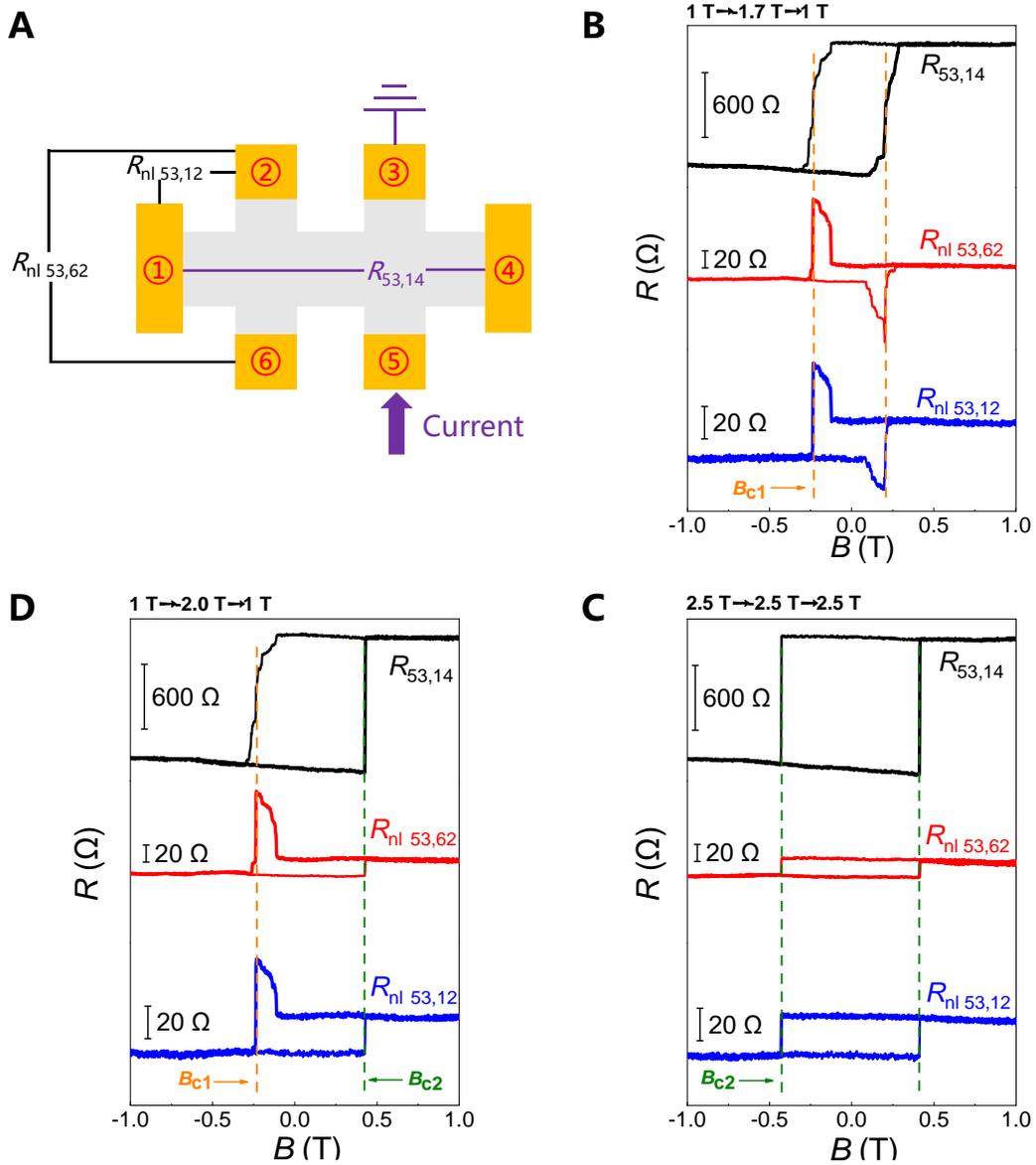

**Fig. 4. The nonlocal measurement of the MnBi$_2$Te$_4$/Cr$_2$Ge$_2$Te$_6$ device.** (**A**) Measurement structure of nonlocal in 5 K and $V_g$ = 60 V. The current is marked by blue arrow and three channels are measured in the meantime. (**B**)-(**D**) The nonlocal transport measurement with two nonlocal channels and one Hall channel. The sweeping range is gradually increased to reveal the variation of $B_{c1}/B_{c2}$ by orange/green marks.